# A versatile quantum microwave photonic signal processing platform based on coincidence window selection technique

Xinghua Li[1,2], Yifan Guo[1,2], Xiao Xiang[1,2], Runai Quan[1,2], Mingtao Cao[1,2], Ruifang Dong[1,2,3*], Tao Liu[1,2,3], Ming Li[4,5,6], Shougang Zhang[1,2,3]

[1]*Key Laboratory of Time Reference and Applications, National Time Service Center, Chinese Academy of Sciences, Xi'an, 710600, China*

[2]*School of Astronomy and Space Science, University of Chinese Academy of Sciences, Beijing, 100049, China*

[3]*Hefei National Laboratory, Hefei 230088, People's Republic of China*

[4]*State Key Laboratory on Integrated Optoelectronics, Institute of Semiconductors, Chinese Academy of Sciences, Beijing, 100083, China*

[5]*School of Electronic, Electrical and Communication Engineering, University of Chinese Academy of Sciences, Beijing 100049, China*

[6]*Center of Materials Science and Optoelectronics Engineering, University of Chinese Academy of Sciences, Beijing 100190, China*

(*[dongruifang@ntsc.ac.cn](mailto:dongruifang@ntsc.ac.cn)*)

**Abstract**

Quantum microwave photonics (QMWP) is an innovative approach that combines energy-time entangled biphoton sources as the optical carrier with time-correlated single-photon detection for high-speed RF signal recovery. This groundbreaking method offers unique advantages such as nonlocal RF signal encoding and robust resistance to dispersion-induced frequency fading. This paper explores the versatility of processing the quantum microwave photonic signal by utilizing coincidence window selection on the biphoton coincidence distribution. The demonstration includes finely-tunable RF phase shifting, flexible multi-tap transversal filtering (with up to 15 taps), and photonically implemented RF mixing, leveraging the nonlocal RF mapping characteristic of QMWP. These accomplishments significantly enhance the capability of microwave photonic systems in processing ultra-weak signals, opening up new possibilities for various applications.

**Introduction**

Microwave photonics (MWP)[1-4] is an interdisciplinary field that combines the principles of both microwave engineering and photonics to leverage the unique advantages of both domains. Due to its wide range of applications across various areas including broadband wireless access networks, sensor networks, radars, satellite communications, and warfare systems[5-11], MWP has been intensively researched for the last few decades and numerous solutions have been demonstrated[12-20]. Accompanying with the rapid development of MWP technology, the bandwidth and sensitivity limitations in traditional microwave boost the highly demanding need for the exploration of new technology to enhance the capability of microwave photonics[21]. Inspired by the enhancements of photonic quantum technology in a vast range of fields from precise navigation and timing, secure communications, to super-resolution imaging and sensing etc.[22-30], QMWP is highly promising to break the bottlenecks of the current microwave photonics technology.

By applying the super low-jitter and high-sensitivity single-photon detector, the scheme of single-photon MWP (SP-MWP) has provided the capability of ultra-weak signal detection and high-speed processing[28]. Further utilizing energy-time entangled biphoton source as the optical carrier, the QMWP technology has been proposed and presented in Radio-over-fiber (ROF) system[29]. Benefitting from the nonclassical feature of quantum entanglement, the unprecedented capability of nonlocal RF signal modulation with strong resistance to the dispersion-induced frequency fading effect associated with ultrashort pulse carriers as well as the significantly improved spurious-free dynamic range (SFDR) in terms of second harmonic distortion have been shown, unveiling invaluable new possibilities in microwave photonics. Recently, both RF phase shifting and multi-tap transversal filtering have been demonstrated by introducing a programmable optical waveshaper into the QMWP system[30]. However, the wavelength selection onto the photon carrier can lead to a significant loss, at least 20 dB for realization of a 3-tap transversal filtering, to the system. Besides, constrained by the grating-based monochromator used in the waveshaper, the achievable spectral resolution is challenging to improve for the purpose of precise and efficient wavelength selection[31]. On the other hand, as the implementation of the QMWP technology depends directly on the coincidence-based heralding on either the signal or the idler photons, the law of appropriate selection of the coincidence window width for maximizing the recovered RF signal has been revealed very recently[32].

In this paper, the utility of the coincidence window selection for processing the RF signal is further revealed, which encompasses the versatile realization of a microwave photonic phase shifter, microwave photonic filter, and a microwave photonic mixer. The phase shifter offers a fine phase shift of 0.01 rad @ 0.2 GHz and a large phase shift range of 24.5 rad @ 6.1 GHz, achieved by varying the window displacement. For multi-tap filters, the tap number, free spectrum range (FSR) and main side lobe suppression ratio (MSLR) are key indicators of their performance, which can be flexibly manipulated by configuring the number of selection windows within the biphotons coincidence distribution envelop, adjusting the spacing between these windows, and applying the required weight ratios to them. Furthermore, we present a novel microwave photonic mixer structure based on the nonlocal RF mapping characteristic of QMWP, which offers enhanced functionality and flexibility in RF signal mixing applications. These accomplishments once again give prominence to the superiority of QMWP and its bright prospect in the new application field of microwave photonics.

## Results

**Theoretical principle** To illustrate the working principle of the QMWP processing scheme based on the coincidence window selection technique, its comparison between the classical MWP scheme is first presented. Fig. 1 (a) demonstrates the implementation of a basic classical multi-tap transversal MWP filter using a multi-wavelength optical carrier. This carrier is modulated by a high-speed RF signal and undergoes dispersion. According to the dispersive phase shifting theory, each wavelength component in the optical carrier experiences a wavelength-dependent phase shift due to dispersion. By setting the wavelengths of the optical carrier as $\lambda_1, \cdots, \lambda_n$ with an identical spacing between them, the n-tap transversal MWP filter is realized[33].

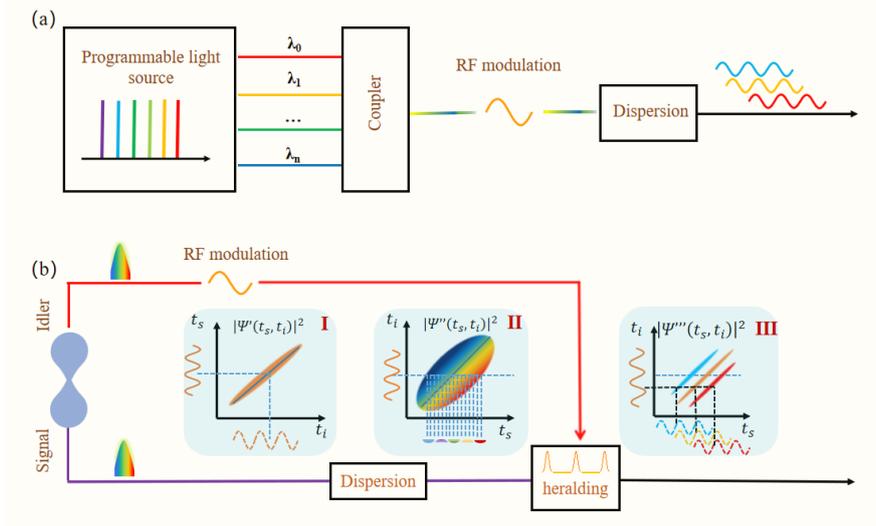

Fig.1 (a) Scheme of a classical n-tap transversal MWP filter. (b) Scheme of a QMWP transversal filter, insets (I) - (III) depict the working principle of the QMWP signal processing based on coincidence window selection technique.

In the QMWP transversal filtering scheme, as shown in Fig. 1 (b), the RF modulation on the idler photons can be considered as the temporal shaping of the photon flows. This shaping is nonlocally mapped onto their "twin" signal photons due to the energy-time entanglement between the signal and idler photons[34-35]. The mapping process is visualized in inset I of Fig. 1 (b). Simultaneously, the dispersion applied to the signal photons not only broadens the biphoton coincidence distribution but also enables nonlocal wavelength-to-time mapping[36]. The joint temporal distribution profile of the dispersed photon pairs is visualized in inset II of Fig. 1 (b). By incorporating multiple selection windows, spaced identically by $\gamma$, within the biphoton coincidence distribution envelope, the QMWP transversal filtering function is implemented. The visualized evolution of this process is displayed in inset III of Fig. 1 (b).

The theoretical principle is described as well. Consider that a continuous wave pumped spontaneous parametric down conversion (SPDC) is utilized to generate the energy-time entangled biphoton source, its two-photon spectral wave function in time domain can be given by[32]

$$\Psi(t_1, t_2) \propto \exp[-i(\omega_{s,0}t_1 + \omega_{i,0}t_2)] \exp\left[-\frac{\sigma^2(t_1-t_2)^2}{2}\right], \quad (1)$$

where $t_1$ and $t_2$ denote the temporal coordinates of the emitted optical signal and idler photons, respectively. $\omega_{s(i),0}$ denotes the center angular frequency of the signal (idler) photon, and $\sigma$ is the correlation time of the twin photons. Then the idler photons are intensity modulated by the RF signal, while the the signal photons pass through a dispersive medium for the (nonlocal) wavelength-to-time mapping. Assume the RF signal has a frequency $\omega_{RF}$, its transfer function is given by $M(t_2) = 1 + A\cos(\omega_{RF}t_2)$, where $0 < A \leq 1$ represents the modulation magnitude. The corresponding transfer function for the dispersive element with a dispersion parameter of $D$ is given by $H(t_1) \propto \frac{1}{\sqrt{|D|}} \exp\left[-i\left(\frac{t_1^2}{2D}\right)\right]$. The resultant biphoton temporal waveform can be expressed as

$$\Psi''(t_1, t_2) \propto \int d\tau \, exp\left[-\frac{\sigma^2(\tau-t_2)^2}{2}\right] \exp\left[-i\frac{(t_1-\tau)^2}{2D}\right](1 + A\cos[\omega_{RF}t_2]), \quad (2)$$

whose square module defines the second-order Glauber correlation function ($G^{(2)}$), i.e., $G^{(2)} \equiv |\Psi(t_1, t_2)|^2$, and is measured by the biphoton coincidence distribution. Based on the detailed deduction of $G^{(2)}$, as provided in the supplementary materials, the temporal density function of the signal and idler photons can be approximately expressed as:

$$\rho_i(t_2) \propto \frac{2\pi}{\sigma}(1 + A\cos[\omega_{RF} t_2]), \tag{3}$$

$$\rho_s(t_1) \propto \frac{2\pi}{\sigma}\left(1 + Ae^{-\frac{\omega_{RF}^2}{4\sigma^2}} e^{-\frac{\sigma^2 D^2 \omega_{RF}^2}{4}} \cos[\omega_{RF} t_1] \cos\left[\frac{D\omega_{RF}^2}{2}\right]\right). \tag{4}$$

From Eq. (3) and (4) one can see that, the RF modulation on the idler photons can be nonlocally mapped onto their entangled counterparts. The dispersion in the signal photon path both degrades the amplitude by a factor of $e^{-\frac{\sigma^2 D^2 \omega_{RF}^2}{4}}$ and introduces the RF-dependent fading ($\cos\left[\frac{D\omega_{RF}^2}{2}\right]$) to the recovered the RF signal. As demonstrated in Ref. [32], by appropriately selecting the width of the coincidence window, which is described by the function $F(t_1 - t_2) = \exp\left[-\frac{(t_1-t_2)^2}{2\alpha^2}\right]$, it is possible to minimize the two adverse effects. In the subsequent discussion, we unveil that a straightforward modification of this function can significantly enhance the feasibility of RF signal processing. The modified function is expressed as follows:

$$F'(t_1 - t_2) = \sum_k \exp\left[-\frac{(t_1 - t_2 - k\gamma)^2}{2\alpha^2}\right], \tag{5}$$

where $k$ is integer and represents the $k$-th coincidence selection window, whose center has a deviation of $k\gamma$ from that of the biphoton correlation distribution. The total number of $k$ which is taken is given by $N$. $\alpha$ and $\gamma$ respectively denote the width of each coincidence selection window and the spacing between them. Assume $\gamma \gg D\omega_{RF}$ and in the approximation of $\alpha^2, \frac{1}{\sigma^2} \ll D^2\sigma^2$, the resultant temporal density function of the signal photons can be deduced as

$$\rho_s'(t_1) = \int dt_2 |\Psi''(t_1, t_2)|^2 F'(t_1 - t_2)$$

$$\propto \frac{2\pi\alpha}{D\sigma^2}\left(1 + Ae^{-\frac{\alpha^2 \omega_{RF}^2}{2D^2\sigma^4}} \cos\left[\frac{\omega_{RF}^2}{2D\sigma^4}\right] \sum_k e^{-\left(\frac{k\gamma}{D\sigma}\right)^2} \cos[\omega_{RF}(t_1 - k\gamma)]\right). \tag{6}$$

For the case of $N = 1$ and $k = \pm 1$, a QMWP phase shifter can be implemented, where the relative phase shift is generated by the displacement between the selection window and the biphoton correlation distribution. The relation between the phase shift and the displacement $\gamma$ can be described as $\varphi = \pm\gamma\omega_{RF}$. By changing the window displacement $\gamma$, the phase shift can be flexibly tuned for a given $\omega_{RF}$. For the condition of $N = 3$ and $k = 0, -1, 1$, a 3-tap QMWP transversal filter is constructed, with the free spectral range (FSR) being $FSR = 1/\gamma$. By increasing the number ($N$) of coincidence selection windows, as long as all of them fall within the biphoton correlation distribution, the tap number of the filter can be flexibly extended.

**Quantum microwave photonic phase shifter** To realize a phase shifter, the nonlocal RF signal mapping on the signal photons and its phase shifting based on the coincidence window selection technique is first investigated. The RF modulation is applied to the idler photons at a frequency of 2.08 GHz with a power of 10 dBm. When the dispersion compensation module (DCM) in the signal photon path is set to have a group delay dispersion (GDD) of 495 ps/nm, the measured full width at half maximum (FWHM) of the biphoton coincidence distribution width is broadened to 300 ps. By applying different window selection of Fig. 2 (a) to the biphoton coincidence distribution, the reconstructed waveforms of the signal and idler photons are plotted in Fig. 2 (b) and 2 (c) respectively. In these plots, we choose three different window displacements ($\gamma$) relative to the center of the biphoton coincidence distribution: 0 ps, 120 ps, and 240 ps. The window width is fixed at 48 ps, achieved by selecting photons within specific time bins of the TCSPC measured histogram, which is comparable to the timing jitter of the SNSPDs. At $\gamma = 0$ ps, both the signal and idler photon waveforms exhibit the same phase. As $\gamma$ increases, it becomes evident that the phase of dispersed signal photons undergoes a right-hand shift in their phase. However, the non-dispersed idler photons remain unchanged regardless of the window displacement. Additionally, it can be observed that the amplitude of the recovered microwave signal gradually decreases with increasing window displacement ($\gamma$). This decrease is attributed to the reduced coincidence counts at positions deviating from the center of the biphoton coincidence distribution. To further clarify this effect, Fig. 2 (e) plots the amplitudes of the recovered RF signals from the signal photon path as a function of the center displacement. We investigate three different modulation frequencies: 0.2 GHz (black squares), 4.1 GHz (orange dots), and 6.1 GHz (red purple triangles). The solid lines represent the theoretical fittings to these results using Eq. (6), and we observe good agreement between the theoretical simulation and the experimental data.

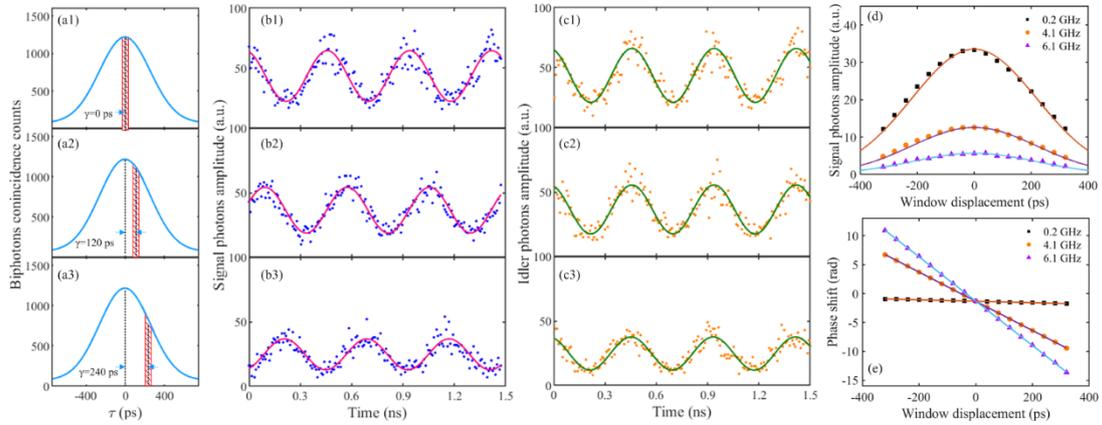

Fig.2 (a) Different window displacements in the biphoton coincidence distribution. (b) - (c) The reconstructed waveforms from the signal and idler photons based on the coincidence window selection technique. Their center deviations are respectively: (b1) - (c1) $\gamma = 0$, (b2) - (c2) $\gamma = 120$ ps, (b3) - (c3) $\gamma = 240$ ps; (d) The experimentally acquired amplitudes of the recovered RF signals from the signal photon path as a function of the center deviation. For all the results, the selected windows have a width of 48 ps. The GDD of the DCM was chosen as 495 ps/nm and the RF modulation is at 2.08 GHz. (e) The extracted phase shift as a function of the window displacement. Three modulation frequencies at 0.2 GHz, 4.1 GHz, and 6.1 GHz are investigated and shown by black squares, orange dots and purple triangles respectively.

To evaluate the phase-shifting performance of the system with different window displacements, we extracted the phase of the signal photon waveforms at various window positions. Fig. 2 (e) shows the phase differences as a function of the window displacement for three modulation frequencies at 0.2 GHz (black squares), 4.1 GHz (orange dots), and 6.1 GHz (red purple triangles). We can clearly observe a linear correlation between the amount of phase shift and the window displacement. This demonstrates the fine-tuning capability of the phase shift for low modulation frequencies and the large dynamic range of the phase shift for high modulation frequencies. In our system, the time-bin resolution of the biphoton coincidence counts distribution is set to 8 ps, resulting in a minimum window displacement $\gamma$ of 8 ps. As a result, at the low modulation frequency of 0.2 GHz, the minimum achievable phase shift is 0.01 rad. On the other hand, to ensure a good recovery of the signal photons' waveforms for phase analysis, the maximum window displacement is limited to the FWHM of the biphoton coincidence distribution. Consequently, at the maximum window displacement, the dynamic range of the phase shift reaches a maximum value of 24.5 rad at the modulation frequency of 6.1 GHz, approximately 7.8π. To further validate this phase shifting phenomenon, the phase differences as a function of the window displacement under the condition of GDD=826 ps/nm and different modulation frequencies are also investigated (see Supplement materials).

**Quantum microwave photonic filter** The multi-tap transversal filtering function using coincidence window selection is then evaluated by inserting a DCM with a GDD of 826 ps/nm in the signal photon path. To demonstrate a three-tap filter, we set three windows with displacements of {-240 ps, 0 ps, 240 ps} and identical widths of 48 ps as the tap. The main to sidelobe ratio (MSLR), which indicates the sidelobe suppression capability of the filter, can be flexibly adjusted by assigning different weights to the windows. With the RF modulation frequency varying from 200 MHz to 8 GHz, the ratios between the amplitudes of the recovered RF waveform with dispersion and that without dispersion are investigated and plotted in Fig. 3 by blue diamonds. For three different weight configurations of the windows: (a1) 0.56:1:0.56, (a2) 0.75:1:0.75, and (a3) 1:1:1, the corresponding MSLR are respectively given by 10.62 dB, 6.29 dB, and 3.95 dB. The solid orange curve is the theoretical simulation result based on Eq. (6), which shows a nice agreement with the experimental results. Furthermore, we examine the dependence of MSLR on the weight ratio, which is plotted in Figure 3 (d). The experimental results align perfectly with the theoretical predictions.

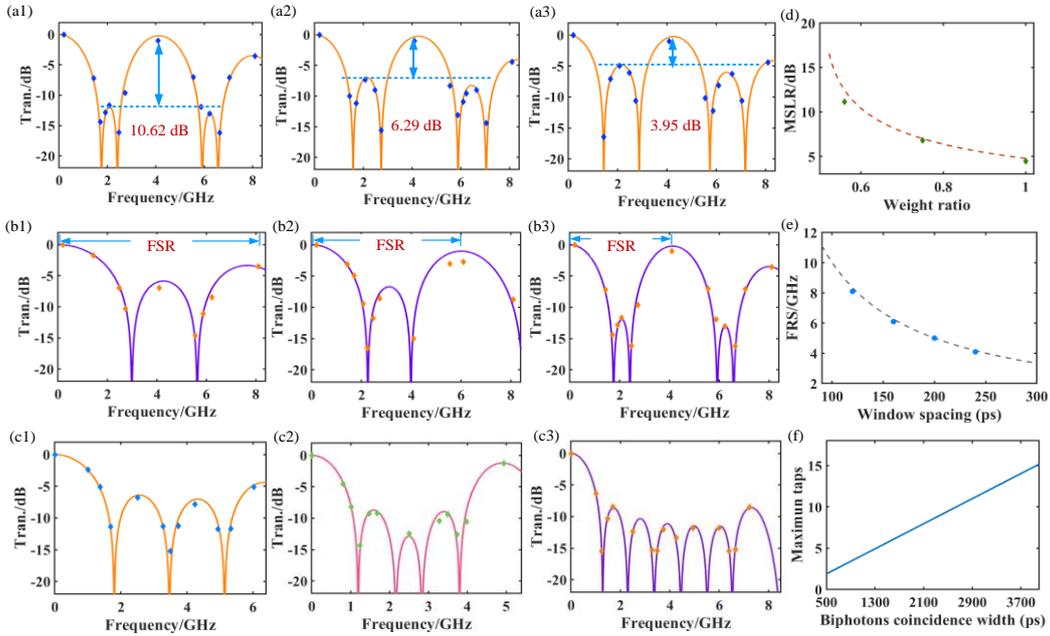

Fig.3 Illustration of the 3-tap transversal filter based on applying three selection windows to the biphoton coincidence measurement. In (a), the displacements of the 3 windows are set as {-240 ps, 0 ps, 240 ps} with their widths being identical as 48 ps. (a1) - (a3) Reconfigurable MSLR of the 3-tap filter is investigated via setting the weights of the three windows at 0.56:1:0.56, 0.75:1:0.75, and 1:1:1 respectively. (b1) - (b3) The FSR tunability of the 3-tap filter is investigated while the spacing between the windows is chosen as 240 ps, 160 ps, and 120ps, the weight of the three windows is fixed at 0.56:1:0.56. (c1)-(c3) Illustration of multi-tap filter by increasing the number of selection windows 4-tap, 5-tap, and 7-tap, respectively. (d) Dependence of the MSLR on the weight ratio. (e) The relationship between FSR and the window spacing value. (f) The theoretically achieved maximum tap number as a function of biphoton coincidence width.

To investigate the FSR tunability of the 3-tap filter, we fix the weight of the three windows at 0.56:1:0.56 and vary the spacing between the windows. Fig. 3 (b1)-(b3) demonstrate the resulting FSR values of 8 GHz, 6 GHz, and 4 GHz, respectively, achieved by choosing window spacings of 240 ps, 160 ps, and 120 ps. The achieved FSR values, plotted in Fig. 3 (e) as blue dots, exhibit an inverse dependence on the spacing ($\gamma$) that aligns perfectly with the theoretical expectation. As the Q factor of a filter is directly related to the number of taps[33], the increase of tap number is another important issue. By increasing the number of selection windows, a multi-tap filter can be realized. Fig. 3 (c1)-(c3) illustrate the reconstructed multi-tap QMWP filters, implemented by introducing four, five, and seven selection windows to the biphoton coincidence distribution. The tap number of the filter can be conveniently adjusted by altering the number of windows. If a DCM with a larger GDD is utilized to broaden the biphoton coincidence width, it would be possible to increase the tap number of the filter. Fig. 3 (f) showcases theoretically achievable maximum number of the filter taps as a function of the biphoton coincidence width. A maximum tap number of 15 can be achieved when the biphoton coincidence distribution width reaches 3700 ps at a GDD value of 1650 ps/nm.

**Quantum microwave photonics mixer** Benefitting from the nonlocal RF mapping characteristic of QMWP, the photonic RF mixing is also realized. For demonstrating the RF mixing, a RF signal ($\omega_{RF1}$) at a frequency of 5 GHz and with a modulation power of 10 dBm is intensity modulated onto the signal photons, whose temporal waveform is Fourier transformed and shown in Fig. 4 (a1). At the same time, a RF signal at a frequency of 1 GHz ($\omega_{RF2}$) and with a modulation power of 10 dBm is intensity modulated onto the idler photons, whose temporal waveform is Fourier transformed and shown in Fig. 4 (a2). Applying coincidence-based post-selection, the Fourier spectra of the reconstructed temporal waveforms from the signal and idler photons are then given in Fig. 4(b1) and (b2). One can see that, the signal and idler photons not only acquire the RF component carried by their twins but also give rise to the sum frequency ($\omega_{RF1} + \omega_{RF2}$) and difference frequency ($\omega_{RF1} - \omega_{RF2}$) of the two RF components. To eliminate the individual RF components from the RF mixing components, a filter is required. Installing a DCM with a GDD of 826 ps/nm in the idler photon path, a 2-tap transversal filter is then designed by utilizing the afore mentioned coincidence window selection technique, whose filtering function is shown in Fig. 4 (c1). After the filtering manipulation, the Fourier spectrum of the idler photon waveform is shown in Fig. 4 (c2), which contains only the difference frequency and sum frequency components. Finally, in Fig. 4 (d1), Chebyshev digital filtering algorithm[37] is employed to realize an output with either sum or difference frequency, as demonstrated in Fig. 4 (d2).

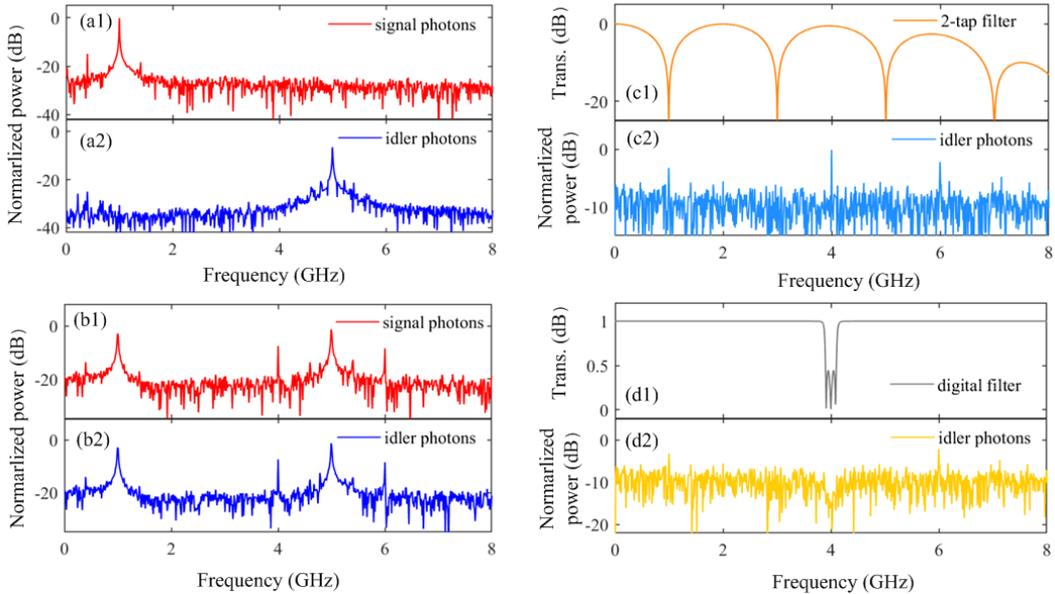

Fig. 4 Fourier spectra of the RF modulated signal photons at 5 GHz (a1) and the RF modulated idler photons at 1 GHz (a2). Fourier spectra of the signal photon waveform (b1) and the idler photon waveform (b2) after the nonlocal RF signal mapping. (c1) The frequency response of the designed two-tap filter. (c2) Fourier spectrum of the idler photon waveform after the two-tap frequency filtering manipulation. (d1) The frequency response of a designed digital filter designed based on the Chebyshev algorithm. (d2) Fourier spectrum of the idler photon

waveform after the digital filtering manipulation.

**Discussion**

The microwave photonic phase shifter developed in this study exhibits exceptional linearity across the entire frequency spectrum, including low, medium, and high frequency bands. This addresses a common issue observed in classical phase shifters, where phase linearity tends to degrade at higher frequencies[38]. Under specific window displacement condition, the maximum achievable phase shift is determined by the detectable microwave signal bandwidth of the system. In our system, the utilized SNSPD has a timing jitter of approximately 50 ps (FWHM). According to the Fourier transformation law, it sets the maximum detection frequency bandwidth to be about 8.8 GHz[28]. Consequently, the calculated maximum phase shift range can reach 18.21π under the condition of a maximum window displacement of 520 ps and a GDD value of 826 ps/nm.

By the same token, the maximum detection frequency bandwidth also determines the maximum FSR that can be obtained by the QMWP filter. As the maximum tunable FSR bandwidth is limited to about 8.8 GHz. it means that the minimum spacing between windows should be larger than 112 ps. In our system, the single photons' bandwidth after SPDC is approximately 2.4 nm. After undergoing dispersion in a DCM with a GDD value of 826 ps/nm, the biphoton coincidence distribution is broadened to having a FWHM of 790 ps. As the tap number cannot be freely increased since all the windows should be located within the FWHM of the biphoton coincidence distribution, the maximum achievable tap number is limited to 7. The current state-of-the-art SNSPDs can achieve a remarkable sub-10 ps timing jitter [40-41]. By utilizing such advanced SNSPDs, the FSR can be increased to 44 GHz, and a maximum tap number of 35 can be achieved with the same dispersion incorporated into the system. As the Q factor is directly related to the number of taps used in its implementation, the Q factor of the QMWP filter can be significantly improved.

The QMWP mixer in our system functions similarly to conventional microwave photonic mixers, where a series connection of two electro-optic intensity modulators is typically employed[19]. However, our system offers enhanced flexibility through the coincidence window selection technique, which eliminates the need to fix the center wavelength of the optical filter. Furthermore, this technique ensures insensitivity to environmental changes while effectively suppressing fundamental frequencies without affecting the output mixing frequency. The conversion loss for this QMWP RF mixer is calculated to be approximately 10-13 dB, which is higher compared with the conversion loss value of classical microwave photonic mixers, typically ranging from 5-10 dB. Despite this higher loss, we are able to achieve a remarkable isolation level of 20 dB, which is defined as the ratio of the leakage power at the output to the input RF signal power. This high level of isolation ensures effective suppression of unwanted signals and minimizes their interference with the desired output.

In conclusion, we have introduced a novel quantum microwave signal processing system that incorporates the coincidence window selection technique. This system seamlessly integrates a microwave photonic phase shifter, microwave photonic filter, and microwave photonic mixer. The phase shifter enables precise and versatile phase shift adjustments through window displacements. By manipulating the ratio of photons within the window and adjusting window spacings, we can achieve a three-tap filter with adjustable MSLRs and tunable FSRs. Compared to previous multi-tap filtering methods that rely on the application of the programmable waveshaper to the photon carrier, our approach offers a cost-effective solution by simply increasing the number of windows within the biphoton wave packet. Additionally, our system features an innovative microwave photonic mixer structure that utilizes nonlocal RF signal mapping and a reconfigurable multi-tap filtering function. This design enhances the functionality and flexibility of signal processing applications. Overall, our proposed system showcases the potential of quantum microwave signal processing techniques and opens up new possibilities for advanced communication systems.

**Materials and methods**

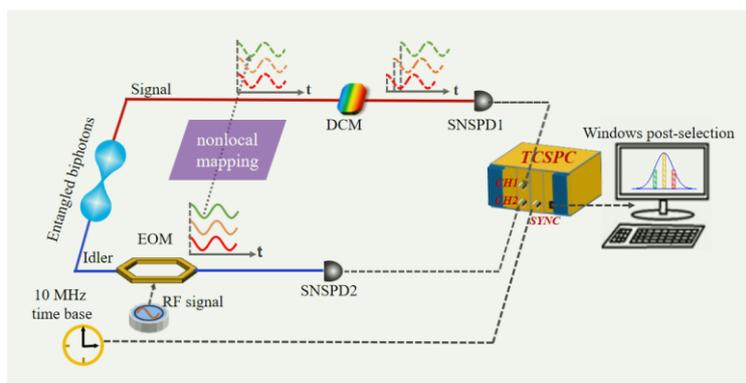

Fig. 5 Experimental setup. EOM: electro-optic modulator; DCM: dispersion compensation module; SNSPD: superconductive nanowire single-photon detector; RF signal: radio frequency signal; TCSPC: time-correlated single photon counting.

The experimental setup of the QMWP signal processing system is shown in Fig. 5. The energy-time entangled photon pairs are generated from a piece of 10 mm long, type-II PPLN waveguide pumped by a CW laser at 780 nm [39]. The idler photons are intensity modulation by a Mach-Zehnder modulator (MZM, PowerBit$^{TM}$ F10-0, Oclaro), through which the high-speed RF signal from a signal generator (E8257D, KeySight) is loaded. The signal photons are fed into a fiber-Bragg-grating-based dispersion compensation module (DCM, Proximion AB) for realizing the dispersion-induced phase shifting. Afterwards, the signal and idler photons are respectively detected by the low-jitter superconductive nanowire single-photon detectors (SNSPD1 & SNSPD2, Photec) with their timing jitter being about 50 ps in full width at half maximum (FWHM, all the mentioned widths in the text refers to the FWHM), which determines the minimum width of the selection window. The two SNSPD outputs are then fed into a Time-Correlated Single Photon Counting (TCSPC) module (PicoQuant Hydraharp 400), which is operated in the Time Tagged Time-Resolved (TTTR) T3 mode with its time-bin resolution being set as 8 ps. The 10 MHz time base from the signal generator E8257D is used for establishing phase stabilization between the RF signal and the sync signal. By manipulating appropriate selection windows on the measured biphoton coincidence distribution, the signal photons can be selected to build the temporal waveforms with desired phase shifting or filtering features.

**Acknowledgments**

This work was supported by the National Natural Science Foundation of China (Grant Nos. 12033007, 61801458, 12103058, 12203058, 12074309, 61875205), the Youth Innovation Promotion Association, CAS (Grant No. 2021408, 2022413, 2023425), the China Postdoctoral Science Foundation (2022M723174).

**Conflict of interest**

The authors declare that they have no conflict of interest.